\documentstyle[12pt,aasms4]{article}

\received{}
\accepted{}
\journalid{}{}
\articleid{}{}

\slugcomment{ to appear in the Astrophysical Journal}

\begin{document}

\def\lsim{\mathrel{\hbox{\rlap{\hbox{\lower4pt\hbox{$\sim$}}}\hbox{$<$}}}}
\def\gsim{\mathrel{\hbox{\rlap{\hbox{\lower4pt\hbox{$\sim$}}}\hbox{$>$}}}}

\title{Cutoff in the TeV Energy Spectrum of Markarian 421 During Strong 
Flares in 2001  }

\author{F. Krennrich\altaffilmark{1}, H.M.~Badran\altaffilmark{2},      
I.H.~Bond\altaffilmark{3}, S.M. Bradbury\altaffilmark{3},
J.H. Buckley\altaffilmark{4},  D.A.~Carter-Lewis\altaffilmark{1},
M.~Catanese\altaffilmark{2},  W.~Cui\altaffilmark{5},  
S.~Dunlea\altaffilmark{6},  D.~Das\altaffilmark{7},
I.~de la Calle Perez\altaffilmark{3}, D.J.~Fegan\altaffilmark{6},
S.J.~Fegan\altaffilmark{2}\altaffilmark{,15} , J.P.~Finley\altaffilmark{5},
J.A.~Gaidos\altaffilmark{5}, K.~Gibbs\altaffilmark{2},
G.H.~Gillanders\altaffilmark{8},  T.A.~Hall\altaffilmark{1},
A.M. Hillas\altaffilmark{3}, J.~Holder\altaffilmark{3}, 
D.~Horan\altaffilmark{6}\altaffilmark{,2},  M.~Jordan\altaffilmark{4}, 
M.~Kertzman\altaffilmark{9}, D.~Kieda\altaffilmark{10},
J.~Kildea\altaffilmark{6}, J.~Knapp\altaffilmark{3},
K.~Kosack\altaffilmark{4},  M.J.~Lang\altaffilmark{8},
S.~LeBohec\altaffilmark{1}, B.~McKernan\altaffilmark{6},
P.~Moriarty\altaffilmark{11},  D. M\"uller\altaffilmark{12},
R.~Ong\altaffilmark{13},   R.~Pallassini\altaffilmark{3},
D.~Petry\altaffilmark{1},  J.~Quinn\altaffilmark{6},
N.W.~Reay\altaffilmark{7}, P.T. Reynolds\altaffilmark{14},           
H.J.~Rose\altaffilmark{3}, G.H. Sembroski\altaffilmark{5},  
R.~Sidwell\altaffilmark{7},  N. Stanton\altaffilmark{7},
S.P.~Swordy\altaffilmark{12}, V.V. Vassiliev\altaffilmark{10},
S.P.~Wakely\altaffilmark{12},  T.C.~Weekes\altaffilmark{2} }

\altaffiltext{1}{Department of Physics and Astronomy, Iowa State
University, Ames, IA 50011}

\altaffiltext{2}{ Fred Lawrence Whipple Observatory, Harvard-Smithsonian
CfA, Amado, AZ 85645}

\altaffiltext{3}{Department of Physics, University of Leeds,
Leeds, LS2 9JT, Yorkshire, England, UK}

\altaffiltext{4}{Department of Physics, Washington University, St.~Louis,
MO 63130}

\altaffiltext{5}{Department of Physics, Purdue University, West
Lafayette, IN 47907}

\altaffiltext{6}{Physics Department, National University of Ireland,
Belfield, Dublin 4, Ireland}

\altaffiltext{7}{Department of Physics, Kansas State University, Manhattan, KS 66506}

\altaffiltext{8}{Physics Department, National University of Ireland,
Galway, Ireland}

\altaffiltext{9}{Physics Department, De Pauw University, Greencastle, 
                   IN, 46135}

\altaffiltext{10}{High Energy Astrophysics Institute, University of Utah,
Salt Lake City, UT 84112}

\altaffiltext{11}{School of Science, Galway-Mayo Institute of Technology,
Galway, Ireland}

\altaffiltext{12}{Enrico Fermi Institute, University of Chicago,   Chicago, IL 60637}

\altaffiltext{13}{Department of Physics, University of California, Los Angeles, CA 90095}

\altaffiltext{14}{Department of Physics, Cork Institute of Technology, Cork, Ireland}

\altaffiltext{15}{Department of Physics, University of Arizona, Tucson, AZ 85721}

\clearpage \begin{abstract} Exceptionally strong and long  lasting flaring
activity of the blazar Markarian~421 (Mrk~421) occurred between January and March 2001.
Based on the excellent signal-to-noise ratio of the data we
derive the energy spectrum between 260~GeV - 17~TeV with unprecedented
statistical precision.  The spectrum is not well described by a simple
power law even with a curvature term.  Instead the data can be
described by a power law with exponential cutoff: $ \rm
{{dN}\over{dE}} \propto \:$ $\rm E^{-2.14 \: \pm \: 0.03_{stat}
\:} \times e^{-E/E_{0}} \: \: m^{-2} \: s^{-1} \: TeV^{-1} $ with $\rm
E_{0} = 4.3 \pm 0.3_{stat} TeV$.  Mrk~421 is the second $\gamma$-ray blazar
that unambiguously exhibits an absorption-like feature in its
spectral energy distribution at 3-6~TeV suggesting that this may be a universal
phenomenon, possibly due to the extragalactic infra-red background radiation.

\end{abstract}


\section{Introduction}


Since the discovery of TeV $\gamma$-rays from the BL Lac objects,
Mrk~421 (Punch et al.\ 1992) and Mrk~501 (Quinn et al.\ 1996),
detailed very high energy observations of these nearby blazars (z = 0.031, z = 0.034)
have been made.  Measurements of flux variation with time,
particularly simultaneous measurements at several wavelengths, constrain models of
particle acceleration and $\gamma$-ray production in the jets.  Spectral energy density
measurements constrain both the models of the jets and of the
infra-red (IR) photon density in the intervening intergalactic medium.
The possibility of absorption of $\gamma$-rays by IR radiation has been 
predicted for some time (see, e.g., Nikishov
1962; Gould \& Schr\`eder 1967; Stecker, De Jager \& Salamon 1992; Biller et al. 1998;
Vassiliev 1999).

The general picture which has emerged for the spectral energy density
of emitted radiation from BL Lacs has two components, a lower one with
energies extended up to about 100 keV attributed to synchrotron
radiation from electrons, and a higher one with energies sometimes extending to the
TeV range, usually attributed to inverse Compton scattering (see,
e.g., Maraschi, Ghisellini \& Celotti 1992, Marscher \& Travis 1996).
There are also competing models (Mannheim \& Biermann 1992; Mannheim 1993; Mannheim 1998)
which assume that the higher energy component arises from protons,
either by proton-induced synchrotron cascades (PIC models) or by decays and/or 
interactions  of secondary particles such as neutral pions and neutrons, or 
synchrotron  radiation from proton beams
(M\"ucke \& Protheroe 2000; Aharonian 2000).  See Catanese and Weekes 
(1999) and Mukherjee (2001) for reviews of TeV observations and an 
overview of relevant models.

Mrk 421 and Mrk 501 are particularly useful in separating
the spectral characteristics intrinsic to the object from absorption
effects in the intervening medium because they have almost the same
redshift.  They also exhibit strong
flares in the TeV energy regime,  well above typical  quiescent levels, 
making detailed spectral measurements possible for both 
(Gaidos et al.\ 1996; Catanese et al.\ 1997; Protheroe et al.\ 1997;
 Aharonian et al. 1997).
  
Measurements by various TeV astronomy groups  have shown that the energy 
spectrum of Mrk~501 is not a simple power law 
(Samuelson et al. 1998; Aharonian et al.\ 1999a;
Djannati-Ata\"{\i} et al.\ 1999; Aharonian et al.\ 2001) but has
significant curvature.  The two-component nature of multiwavelength
blazar spectra implies that, over a sufficiently wide energy range,
TeV spectra must be intrinsically curved.   The measured curvature 
however depends on the distance of the energy range of the data from the IC peak.  
During the strong flaring activity the synchrotron peak of Mrk~501
appears to  shift to above 100~keV (Catanese et al. 1997; Pian
et al. 1998), with the IC peak shifting to several hundred GeV (Samuelson
et al. 1998).  Measurements of the HEGRA collaboration have the
highest energies extending to $\approx $~20~TeV; their spectrum is fit
better with an exponential cutoff at $\approx 6-8$~TeV (Aharonian et
al. 1999a; Aharonian et al. 2001), rather than a simple parabolic
correction to the power law as used in Samuelson et al.  (1998).

Several groups have determined energy spectra for Mrk 421, both
at low average flux levels ($<1$ Crab) (Aharonian et al. 1999b; Krawczynski et al.
2001; Bazer-Bachi et al. 2001) and from intense flares (2.8 - 7.4 Crab) (Zweerink et
al.\ 1997; Krennrich et al.\ 1999a).  Analysis of the intense flare data
showed that Mrk~421 had a  spectral index different from
Mrk~501.  The data could be acceptably fit with a simple
power law, although there was weak evidence for curvature (Krennrich
et al.\ 1999a).   The shape of the spectral energy distribution for 
Mrk~421 (Aharonian et al. 1999b; Krennrich at al 1999b) and
Mrk~501 generally appears independent of the flux level  (Aharonian et al. 1999a),   
although some evidence for spectral variability has  been reported by  
Djannati-Ata\"{\i} et al. (1999) and Krawczynski et al. (2001) for Mrk~501.  

In this Letter, we present results from $\gamma$-ray observations of
Mrk~421 taken during intense flares in January - March 2001
with the Whipple Observatory 10~m telescope yielding a spectrum
spanning the energy range between 260~GeV and 17~TeV.  The spectrum has
high statistical precision and shows a distinct cutoff
with a characteristic energy of about 3-6 TeV.

\section{Observations \& Data Analysis}

The observations were made with the Whipple Observatory 10~m $\gamma$-ray
  telescope equipped with the GRANITE-III high resolution camera
(Finley et al. 2001).  The fine granularity ($\rm 0.12^{\circ}$) of
the 379 photomultiplier camera provides good sensitivity for point
sources.  The sensitive energy range of the instrument is
 $\rm   \approx $~200~GeV  to  greater than 20~TeV. 
Based on finer sampling of $\gamma$-ray images, the linear
      response of the telescope at the highest energies 
      is improved in comparison with previous camera configurations. 

  The use of a different type of photomultiplier
(Hamamatsu R-960, previously Hamamatsu R1398), a complete re-coating
of the mirrors and the installation of a new set of light concentrators 
 necessitated a  comprehensive and detailed new calibration of 
the telescope.
Three methods were used: the first was based on laboratory measurements of
the individual instrument components, the second utilized the calibrated Cherenkov 
light signal from single secondary cosmic-ray muons, and the third used simulated cosmic-ray 
showers to match observed distributions of the parameters of the background images. 
All three methods provide an energy calibration consistent within 20\% for the absolute 
energy  of primary $\gamma$-rays.

The calibration can be evaluated by checking the 
 energy spectrum of the Crab Nebula, which is a standard candle for
TeV $\gamma$-ray astronomy.  The measurements of the spectrum of the
Crab Nebula with the Whipple telescope over several years data 
using different camera configurations,  are all statistically 
consistent,  showing the same flux constant and energy spectral index.  
Therefore, 
we also show the Crab spectrum in this paper for comparison with the 
Mrk~421 spectrum.  The Crab data consist of 15.4 hours of 
on-source observations (zenith angle $\rm < 35^{\circ}$) with the 
same amount of data for the background estimate.

Mrk~421 was more active in 2000 and 2001 than in previous years
of observations, with the most intense flaring episodes in 2001.  The unusually
high flaring states in 2001 provide remarkable statistics, and we have
chosen the strongest flares for the analysis of the spectral energy
distribution. 
Observations    of    flaring states  on    January~21, January~31,  February~1-3,
February~27, March~19,  March~21,  March~26-27, 2001  provide together
$\rm  \approx 23,000$ photons above  260~GeV. 
To test the validity of combining various flares we have also
divided the data into two subsets and found no statistical inconsistency
in the derived energy spectral indices.
 The statistical significance 
of the excess events corresponds  to more than 86 standard deviations.

The selected 2001 flaring data consist of 30.8~hours of on-source
observations at zenith angles less than $\rm 35^{\circ}$.  For 
background comparison, the same amount of off-source data is used from
observations made at similar zenith angles.  The on-source data provide
an average rate of 12.5 $\rm \gamma$/min, corresponding to 3.7 Crab.
The excellent signal to background ratio can be seen from figure~1.
This plot shows the alpha distribution (for explanation of alpha 
parameter see caption of figure 1 or Reynolds et al. 1993) of events  
with energies greater than 1~TeV from the source (solid line) after 
applying $\gamma$-ray  selection criteria.  The  dotted line shows the 
corresponding  distribution for the background measurements.

The data analysis, $\gamma$-ray selection and energy estimate
use the methods developed by Mohanty et al. (1998).  These
$\gamma$-ray selection criteria are derived from parameter
distributions of simulated $\gamma$-ray showers as a function of their
total light intensity (size) in the camera.      The criteria vary with 
size and are set so that they keep  90\% of the $\gamma$-ray images whose 
centroids lie within $\rm  0.4^{\circ} - 1.0^{\circ}$   of the center 
of the camera.      To avoid the difficulties of modeling the trigger 
electronics  we apply an additional cut, requiring that a signal of at least
15.1 photo electrons (p.e.),  13.6 p.e. and 12.1 p.e.  be  present in the three
highest image pixels, respectively.

\section{Results}

The differential energy flux values derived from the intense flaring 
states of Mrk~421 in 2001,  and for comparison the Crab
Nebula,  are shown in figure 2.  The spectrum of the Crab
Nebula can be well fit by a power law of the form:

\smallskip

$ \rm  {{dN}\over{dE}} = (3.11 \pm 0.3_{stat} \pm 0.6_{syst})
\times \: 10^{-7} \: E^{-2.74 \: \pm \: 0.08_{stat}
\:  \pm \:  0.05_{syst}}  \:  \: m^{-2}  \: s^{-1}   \: TeV^{-1} $

\smallskip

\noindent giving a $\chi^2$ of 6.9  for 10 degrees of freedom (probability
73\%).   This result is consistent, within its limited statistics,
 with previous measurements made by the Whipple collaboration 
(Hillas et al. 1998; Mohanty et al. 1998; Krennrich et al. 1999a),     
showing that the analysis methods, the  calibration of the   detector, and 
the reconstruction of energy spectra are  all consistent at the 2 sigma level
including systematic uncertainties.   The first set of errors on the measured spectral 
index are statistical  and the second are systematic.  The systematic errors 
are determined  by varying the gain of the system by 20\%, and by varying 
the cut efficiencies (see Mohanty et al. 1998).

 Fitting a power law to the energy distribution of
Mrk~421 yields the following result:

\smallskip

$ \rm    {{dN}\over{dE}}  \propto   \: E^{-2.64 \: \pm \: 0.01_{stat}
\:  \pm \:  0.05_{syst}}  \:  \: m^{-2}  \: s^{-1}   \: TeV^{-1} $

\smallskip

\noindent giving a $\chi^2$ of 410.7 for 10 degrees of freedom.  The
spectrum is clearly not compatible with a simple power law form.
A curvature fit for the Mrk~421 spectrum  yields:

\smallskip

$ \rm    {{dN}\over{dE}}  \propto    \:$  $\rm  E^{-2.47
\: \pm \: 0.02_{stat} \pm \: 0.05_{syst} \: -(0.51 \pm 0.03_{stat} \pm 0.05_{syst})log_{10}(E) } \:
\: m^{-2} \: s^{-1} \: TeV^{-1} $.

\smallskip

\noindent giving a $\chi^2 = 56.5$ for 9 degrees of freedom with a
chance probability of $\rm 6 \times 10^{-9}$.  As shown in figure 2,
the Mrk~421 spectrum exhibits clear curvature, but a parabola is 
not a suitable shape.  Assuming that part of the curvature in the spectrum is due
to a cutoff, the data can be fit by:

\smallskip

$ \rm    {{dN}\over{dE}}  \propto    \:$  $\rm  E^{-2.14
\: \pm \: 0.03_{stat}  \:}  \times  e^{-E/E_{0}} \:
\: m^{-2} \: s^{-1} \: TeV^{-1} $

\noindent with $\rm E_{0} = 4.3 \: \pm 0.3_{stat} \: (-1.4 \: +1.7)_{syst} \: TeV$.

\smallskip

\noindent While the $\chi^2$ (25.2 for nine degrees of freedom or a
probability of 0.3\%) is not particularly good,    the  fit 
using an exponential cutoff is much better than the fits using
 the two simple forms considered above.

We should point out here that for the first time statistical
errors of the flux measurements in TeV astronomy reached the level
of 2\%, likely making systematic uncertainties dominant.
 Because of this, indicated confidence levels should be accepted with a caution
as they are based only on statistics of detected photons. We have also
examined the flux values at 5.6 and 8.2 TeV, which appear high
to the eye, $\rm 2.6 \sigma_{stat}$ and $\rm 1.8 \sigma_{stat}$, respectively
above the exponential cutoff fit. Considering  the statistical 
uncertainties we do not regard this as a significant feature in the data.
In future, a more complete analysis will
be pursued to understand systematic errors at the percent level.
In derivation of the cutoff energy we have included two most
prominent sources of systematics: uncertainties in the
reconstruction of the spectrum with a cutoff derived from simulation
tests (test spectra), and uncertainty in absolute energy calibration of the
instrument (20\%).

We also tried a "super-exponential" form (e.g., see Stecker, De Jager \& Salamon 1992;
Aharonian et al. 2001),  but with no improvement in quality of fit. The result is: 

\smallskip

$ \rm    {{dN}\over{dE}}  \propto    \:$  $\rm  E^{-2.24
\: \pm \: 0.09_{stat} }  \times  e^{-(E/E_{0})^{1.3\pm0.4}} \:
\: m^{-2} \: s^{-1} \: TeV^{-1} $

\noindent with $\rm E_{0} = 5.8 \pm 1.5_{stat} TeV$.  The $\chi^2$ of the fit is 
24.0 for eight degrees of freedom (probability of 0.2\%).

\smallskip

\section{Discussion}

For the first time, the  Mrk~421 energy spectrum at  very high energy 
has been determined over a sufficiently wide energy range, 
and with sufficient statistical precision, that an important feature (a cutoff)
could be discerned.  The spectrum  is best described by a power law attenuated by an exponential 
cutoff at an energy of $\rm E_{0} =  4.3 \: \pm 0.3_{stat} (-1.4 + 1.7)_{syst} $~TeV. 

The cutoff feature could have several origins, e.g., the termination
of the particle energy distribution in the primary beam, absorption
near the $\gamma$-ray source (Dermer \& Schlickeiser 1994) or
absorption in intergalactic space by IR background radiation
fields.  If the    cutoff energy varied in time, or differed for 
Mrk~421 and 501,
it would not be due to extragalactic absorption.  We examined our
previous published flare spectra of Mrk~421 (Krennrich et al.
1999a; Zweerink et al. 1997) and found them consistent with the present
spectrum.  Fitting the previous Mrk~421 spectrum (Krennrich et al. 1999a) 
with the  parametrization of  
$ \rm {{dN}\over{dE}} \propto \:$ $\rm E^{-2.14 \: \pm \: 0.03_{stat}
\:} \times e^{-E/4.3\:TeV} \: \: m^{-2} \: s^{-1} \: TeV^{-1} $ results
in an acceptable $\chi^2$ of 19.4 for 13 degrees of freedom (P = 11\%).

We also examined our previously published spectrum for
Mrk 501 (Samuelson et al., 1998) and find that its spectrum
can be fit by  $ \rm {{dN}\over{dE}} \propto \: E^{-1.95 \: \pm \: 0.07_{stat}
\:} \times e^{-E/E_{0}} \: \: m^{-2} \: s^{-1} \: TeV^{-1} $ with 
$\rm E_{0} = 4.6  \pm  0.8_{stat} \:  TeV$.   However, fitting the Mrk 501
spectrum with 
$ \rm {{dN}\over{dE}} \propto \:$ $\rm E^{-2.14 \: \pm \: 0.03_{stat}
\:} \times e^{-E/4.3 \: TeV} \: \: m^{-2} \: s^{-1} \: TeV^{-1} $
results in a poor fit with a $\chi^2$ of 37.6 for 13 degrees of freedom 
(P = $\rm 3 \times 10^{-4}$).
The cutoff energy for Mrk~501 is consistent with the cutoff
energy for Mrk~421, but a different spectral index is required for an
acceptable fit.  Thus, we find no inconsistency with the assumption
that the cutoff might be due to the IR background.   More detailed studies 
of spectral variability of Mrk~421 are in progress and will be presented elsewhere. 

Energy spectra for Mrk 421 reported by the HEGRA collaboration
(Aharonian et al. 1999b) are based on data  taken in 1997 and 1998  with
the source at a lower average flux level ($\rm \approx 0.5$~ Crab).  However,
the primary sensitivity of the HEGRA array is at energies above 1~TeV
requiring a comparison above that energy.  Those results indicate a
power law spectrum of $ \rm {{dN}\over{dE}} \propto \: E^{-3.09 \: \pm
\: 0.07 \pm \: 0.1 }$.  When the Mrk 421 spectral data presented
here are fit with a power law using only data points above 1~TeV, a
spectral index of  $\rm -3.01 \: \pm \: 0.03_{stat} \pm \: 0.05_{syst} $ 
with a $\chi^2$ of 106.8 for 6 degrees of freedom is measured
 ($\rm -2.73 \: \pm \: 0.04_{stat} \pm \: 0.05_{syst} $  if the 1.2~TeV - 8.2~TeV 
range is fitted, $\chi^2$ of 11.1 for 4 degrees of freedom) which is consistent 
with the HEGRA result, considering statistical and systematic uncertainties 
between two different instruments.

In summary, we have presented a Mrk~421 spectrum over the energy
range 260~GeV to 17 TeV which shows an exponential-like cutoff in
the range 3-6 TeV, similar to the cutoff found earlier for Mrk~501
(see figure 3) which has almost the same redshift.  The new Mrk~421
spectrum is consistent with those reported earlier both by our group
and others.  
The occurrence of a cutoff at similar energy 
 in the very high energy spectra of the two equally distant  $\gamma$-ray blazars
Mrk~421 and Mrk~501 might indicate that TeV $\gamma$-rays are absorbed by 
IR background  radiation fields.   Future investigations regarding the implications 
for the energy density of extragalactic IR background radiation  are underway 
(Vassiliev 2001).

\acknowledgments
The VERITAS Collaboration is supported by the U.S. Dept. of Energy,
N.S.F., the Smithsonian Institution, P.P.A.R.C. (U.K.) and
Enterprise Ireland.

\vfill\eject

\clearpage

\begin{center}
 
\mbox{\epsfysize=0.60\textheight\epsfbox{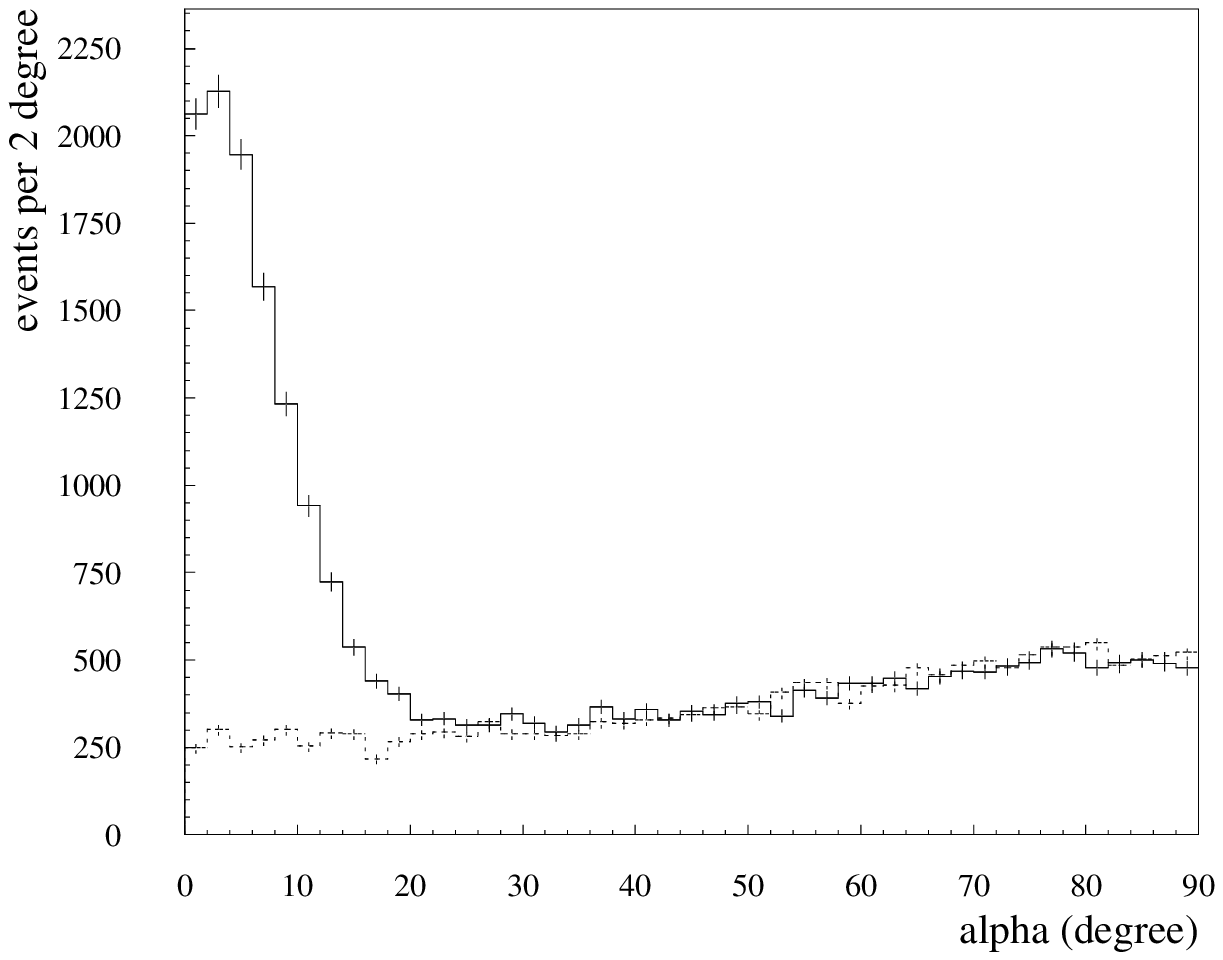}}
  \end{center}

  \figcaption[m42d]
  {The alpha distribution for on-source events (above $\rm \approx$~1~TeV) 
 shown by the solid line and background events (off-source, dashed line) 
    for the 30.8~hours of observations of flaring states of Mrk~421.   The orientation angle alpha
    is defined as the angle between the major axis of an elliptical $\gamma$-ray image and the source
    position.    }

\clearpage

\begin{center}
 
\mbox{\epsfysize=0.60\textheight\epsfbox{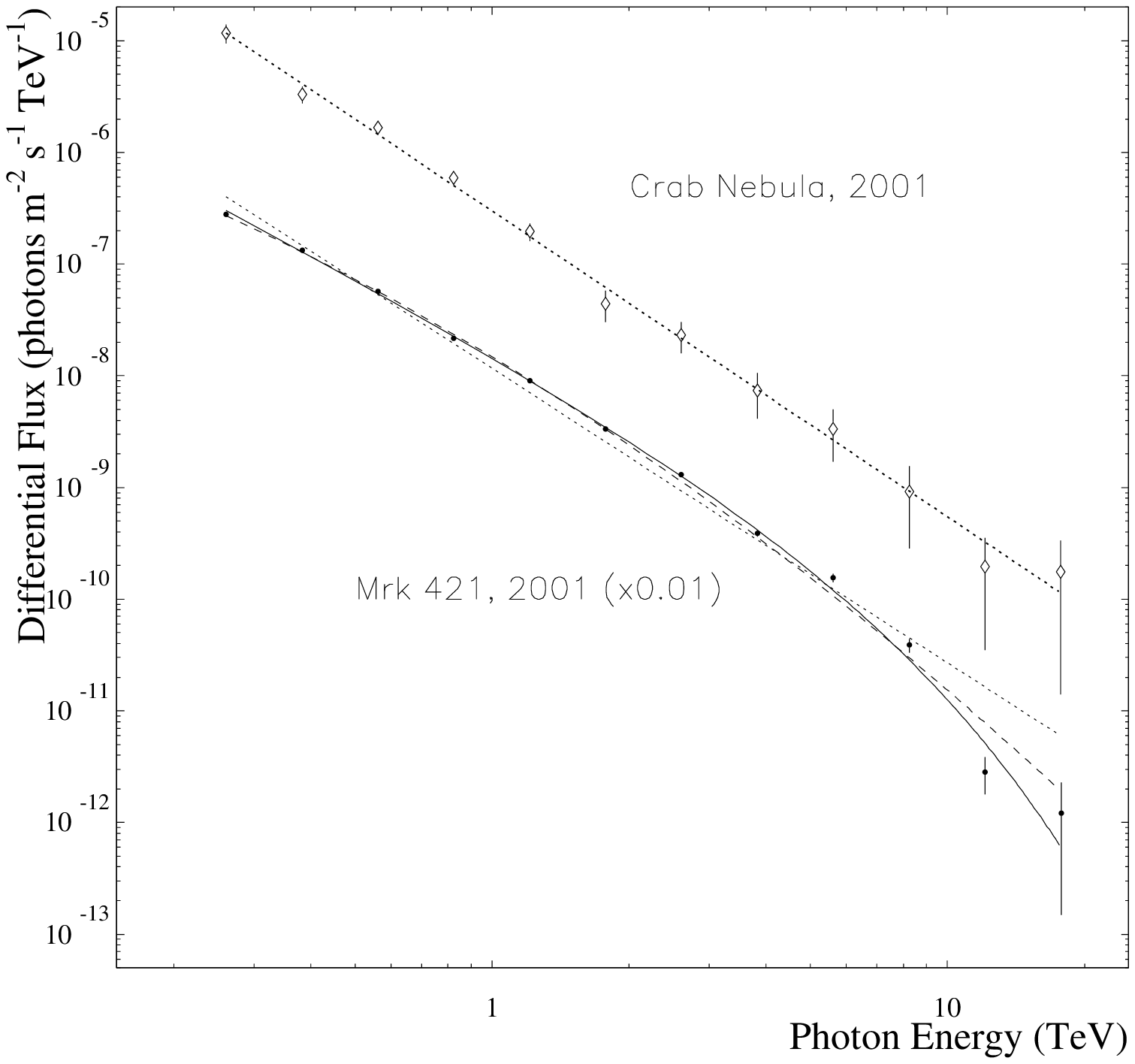}}
  \end{center}

  \figcaption[m42d]
  {The energy spectra of Mrk~421 (filled circles) and the Crab Nebula
(diamonds) for the 2001 data set are shown.  The dotted  lines correspond 
to power law fits, the dashed line
(Mrk~421) corresponds to a parabolic fit, and the solid line is the result from a
fit with an exponential cutoff.  Note that the Mrk~421 spectrum has been offset 
by a factor of 0.01 in flux for clearer presentation, and errors shown include only statistical 
uncertainty.  }

\clearpage

\begin{center}
 
\mbox{\epsfysize=0.60\textheight\epsfbox{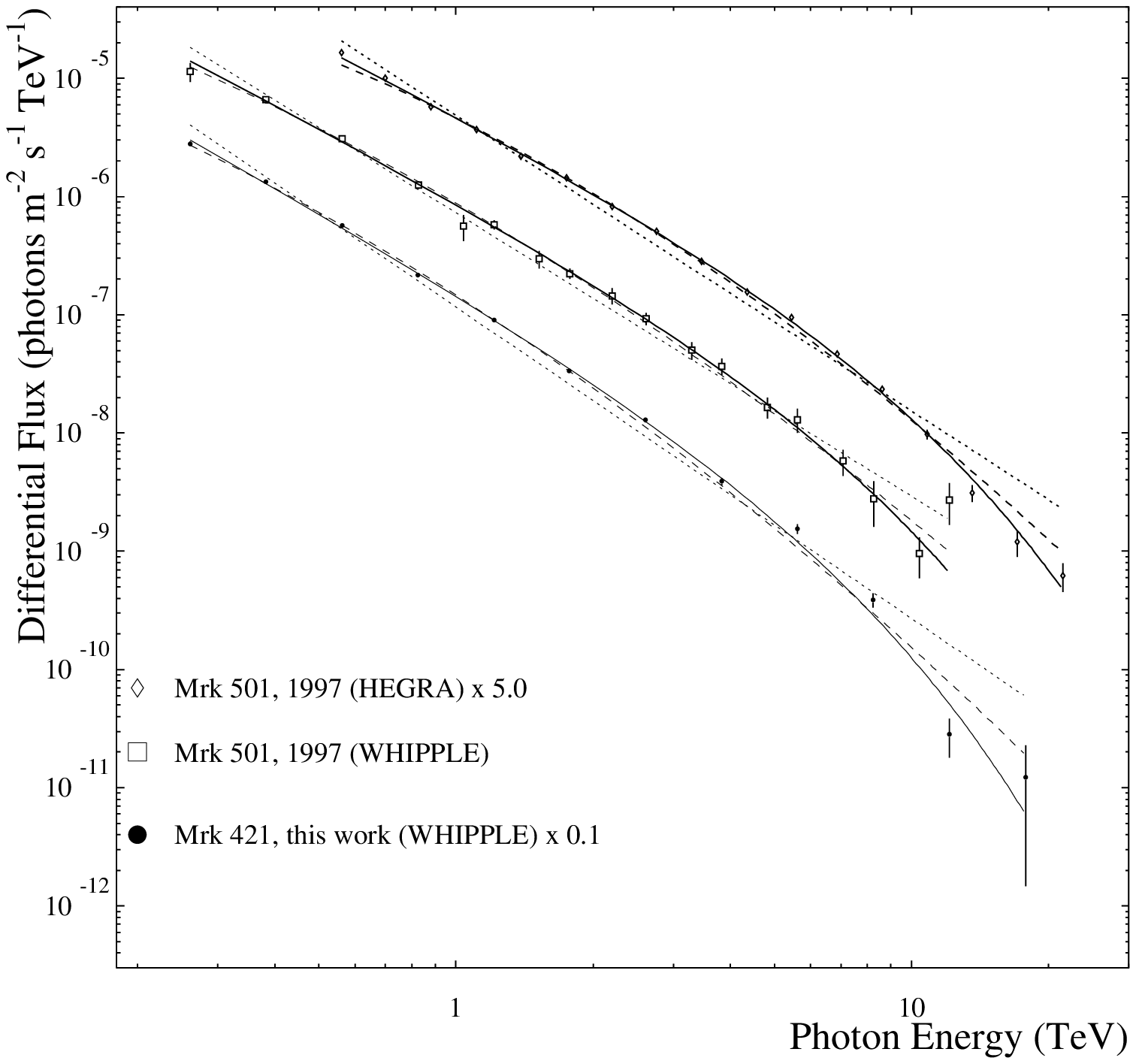}}
  \end{center}

  \figcaption[m42d]
  {A comparison of the Mrk~501 spectra published by the Whipple (Samuelson et al. 1998)
 and the HEGRA  collaboration (Aharonian et al. 1999a) and the Mrk~421 spectrum (this work)
 is shown.    The dotted lines 
indicate power law fits, the dashed lines are parabolic fits and the solid lines show
the power law plus exponential cutoff fits.  A common feature to all three spectra is 
a cutoff at 4 - 6 TeV.  Also note that both the Mrk~501 spectrum (Aharonian et al. 1999a)
and the Mrk~421 spectrum reported in this paper are not well described by a parabolic spectrum.
Also note that spectra have been offset for clearer presentation. }

\end{document}